\documentclass[10pt, fullpage, lefteqn]{scrartcl}  
\date{}
\usepackage{amsmath}
\usepackage{amsfonts}
\usepackage{authblk}
\usepackage{subfig}
\usepackage{graphicx} 
\usepackage{graphics}
\usepackage{times}
\begin{document}
\title{The contributions of surface charge and geometry to
  protein-solvent interaction}
\subtitle{Soluble proteins are capacitors with net negative charge}
 \author{Lincong
  Wang\footnote{Corresponding author: Lincong Wang, Email: {\tt
      wlincong@hotmail.com.}} } \affil{\small The College of Computer
  Science and Technology, Jilin University, Changchun, Jilin, China}

\maketitle

\begin{abstract}
To better understand protein-solvent interaction we have analyzed a
variety of physical and geometrical properties of the solvent-excluded
surfaces (SESs) over a large set of soluble proteins with crystal
structures. We discover that all have net negative surface charges and
permanent electric dipoles. Moreover both SES area and surface charge
as well as several physical and geometrical properties defined by them
change with protein size via well-fitted power laws. The relevance to
protein-solvent interaction of these physical and geometrical
properties is supported by strong correlations between them and known
hydrophobicity scales and by their large changes upon protein
unfolding. The universal existence of negative surface charge and
dipole, the characteristic surface geometry and power laws reveal
fundamental but distinct roles of surface charge and SES in
protein-solvent interaction and make it possible to describe solvation
and hydrophobic effect using theories on anion solute in protic
solvent. In particular the great significance of surface charge for
protein-solvent interaction suggests that a change of perception may
be needed since from solvation perspective folding into a native state
is to optimize surface negative charge rather than to minimize the
hydrophobic surface area.
\end{abstract}

\section{Introduction}
The quantification of protein-solvent interaction is essential for
understanding protein folding, stability, solubility and
function. Along with experimental studies considerable efforts have
been made to quantify protein-solvent interaction using structures
ever since the publication of the first protein crystal structure more
than 50 years ago~\cite{Kendrew1960}. The solvent excluded surface
(SES)\footnote{Abbreviations used:SES, solvent excluded surface; SEA,
  solvent excluded area; SA, solvent accessible; ASA, accessible
  solvent surface area; SAA, solvent accessible area ($a_{s}$); PAA,
  probe accessible area ($a_{p}$); TAA, torus accessible area
  ($a_{t}$); VDW, van der Waals; 2D, two-dimensional; 3D,
  three-dimensional; MD, molecular simulation; BSP, binary space
  partition; SSE, Streaming SIMD Extensions.}~\cite{Lee1971379} of a
protein is a two-dimensional (2D) manifold that demarcates a boundary
between the protein and its solvent.  An SES consists of three
different types of area: solvent accessible area (SAA, $a_s$), torus
accessible area (TAA, $a_t$) and probe accessible area (PAA,
$a_p$). Geometrically $a_{s}$ is a convex area, $a_{p}$ a concave one
and $a_{t}$ a saddle area. Either SES or more frequently accessible
solvent surface~\cite{ASAdefinition, FMRichards1977} has been studied
extensively for its role in protein-solvent
interaction~\cite{Chothia1974, Tanford1974, FMRichards1977,
  Eisenberg1986, OOI1987, Sharp106, charmm2009} since they could be
computed readily using atomic coordinates and radii.  Though much
progress has been made in the past~\cite{CONNOLLY:PQMS, MSMS} there is
still a lack of accurate and robust algorithm and/or efficient
implementation for SES computation and previous SES applications to
protein-solvent interaction have been limited in several respects. Two
most popular programs, PQMS~\cite{CONNOLLY:PQMS} and MSMS~\cite{MSMS},
are neither robust nor accurate enough for a large-scale
application. For example, both may require modification to atomic
radii to handle singular cases of intersecting PAAs and tend to fail
on large proteins with $>$10,000 atoms. A manual restart is needed for
MSMS to compute each internal cavity inside a protein. The arbitrary
modification to radii and manual restart introduce inconsistency and
make them less suitable for accurate SES computation on a large scale.
To overcome these difficulties we have developed a robust SES
algorithm that treats all the possible probe intersecting cases
without any modification to atomic radius. It computes the internal
cavity automatically with no manual intervention and achieves a high
precision with an estimated error $<3.0\times 10^{-3}$\AA$^2$~per
surface atom for all the three types of SES areas\footnote{The total
  error $E_r$ for a protein increases slowly with the number of atoms
  $N$: $E_r \propto \sqrt{N}$}. In our implementation any atom with
$a_s > 3.0\times 10^{-4}$\AA$^2$~is defined as a surface atom.
Previous SES applications to protein-solvent interaction have focused
more on (1) accessible solvent surface area (ASA)~\cite{CHOTHIA19761,
  Sharp106, Ashbaugh1999, charmm2009} rather than solvent-excluded
surface area (SEA) likely due to the easy computation of the former,
(2) individual residues rather than individual atoms and (3) surface
area only rather than the geometrical and physical properties of the
surface. A few studies~\cite{Singh2006145, Kapcha2014484} have used
atomic ASAs for example to predict solvent accessibility of amino acid
residues. To our knowledge no atomic SES applications have been
published up to now. The accuracy and robustness of our algorithm
enable us to perform a statistical analysis on SES's contribution to
protein-solvent interaction over a set $\mathbb N$ of $24,024$ soluble
proteins with crystal structures by focusing on their individual atoms
and using not only SEA but also SES's geometrical and physical
properties especially electrical properties. We discover that not only
every structure in $\mathbb N$ has a net negative surface charge and
permanent electric dipole but the changes with protein size of surface
charge, dipole and surface geometry as well as several physical and
geometrical properties defined by them also follow well-fitted power
laws\footnote{In this paper a linear equation $y=ax+c$ is treated as a
  special power law: $y=ax^b+c$ with $b = 1$.}. For example, the
charge per atom for all the surface atoms has an average of
$-29.6\times 10^{-3}$ coulomb over all the structures in $\mathbb N$
while the charge per atom for all the atoms has an average of only
$-1.6\times 10^{-3}$ coulomb.  Thus soluble proteins behave like an
electric dipole or more properly a capacitor in solution. Moreover our
analysis shows that on average the SEAs for hydrophilic atoms which
are capable of forming a hydrogen bond with a solvent molecule are
almost 2-fold larger than those for hydrophobic atoms. The larger the
SEA of a surface atom is the better of its hydrogen bonding
interaction with solvent. Geometrically we find that concave-convex
ratio $r_{cc} = \frac{a_p}{a_s}$ increases with protein size and upon
folding. The larger of $r_{cc}$ of a surface atom is the flatter of
its local surface.  Most interestingly hydrophobic atoms have larger
$r_{cc}$s than hydrophilic ones. One plausible explanation is that for
a surface atom the larger its $r_{cc}$ the better of its van der Waals
(VDW) interaction with the solvent~\cite{}.  The relevance of these
physical and geometrical properties to protein-solvent interaction is
collaborated by the strong correlations between their values computed
for individual amino acid residues and five well-known hydrophobicity
scales~\cite{Janin1979, KYTE1982105,EWscale1984,GES1986} and by their
large changes upon protein unfolding. For example, the fitted
solvation parameters ($\sigma_i$s)~\cite{Eisenberg1986, OOI1987} in
solvation free energy-surface area relation, $\Delta G_\mathrm{solv} =
\sum_{i} \sigma_{i} \ ASA_{i}$ where $ASA_{i}$ is the ASA for residue
$i$, could be interpreted as surface charges. Previously no physical
meanings have been given to these $\sigma_i$s. In addition our
large-scale analyses of protein-ligand interfaces~\cite{sesProtLigand}
where the ligand is either a DNA or a small-molecule compound or
another protein show that (1) the values for these properties over the
set of surface atoms that are buried upon ligand-binding differ
largely from those over the set of atoms that remain exposed, and (2)
ligand binding share many similarities with protein folding in terms
of SES's physical and geometrical properties.  \\

How a protein interacts with solvent is of paramount importance for
understanding protein folding.  For example, it is widely accepted
that hydrophobic effect is the driving force for protein
folding~\cite{Kauzmann1987, Dill1990}.  However, the nature of
hydrophobic effect and the details of protein hydration shells remain
unclear at present~\cite{Baldwin2014}. The findings presented here
reveal fundamental but distinct roles of surface charge, hydrogen
bonding and SES geometry for protein-solvent interaction and are
consistent with water being a protic solvent that prefers anions over
cations as its solutes. They shed new lights on hydrophobic effect by
demonstrating that it is an effect to which both surface area and
charge contribute and suggest that the optimization of protein
solvation through natural selection is achieved by (1) universal
enrichment of surface negative charge, (2) increased surface areas for
hydrophilic atoms for better hydrogen bonding with solvent, and (3)
higher concave-convex ratio for hydrophobic atoms for better VDW
attraction with solvent. It seems to us that a paradigm shift may be
required in the study of the protein-folding problem by focusing on
surface charge rather than side chain hydrophobicity since folding
into a native state is to maximize the negative surface charge rather
than to minimize the hydrophobic surface area. The statistical values
for the surface charge and dipole moment and the fitted power laws
obtained on the large set of structures should be useful for the
quantification of solvation and hydrophobic effect using well-known
theories on anion solutes in protic solvent~\cite{Onsager1936,
  CKmodel1957, WARSHEL1976227, Warshel20061647,
  solvationModel2009}. Furthermore the statistical values for
SES-defined physical and geometrical properties could also be used to
restraint the folding space for a protein or serve as a term in an
empirical scoring function for protein structure
prediction~\cite{PROT:PROT21715} or a quantity for quality control in
structure determination~\cite{Kota15082011}.  {\em The relative
  importance of surface dipole, hydrogen-bonding and VDW attraction
  for protein-solvent interaction.}

\section{Materials and Methods}
In this section we first describe the data sets used in our
statistical analysis and then briefly present our algorithm for SES
computation. Finally we define a variety of SES-defined physical and
geometrical properties for a protein that are relevant to its
interaction with solvent.
\subsection{The protein data sets}
We have downloaded from the current version of the PDB a non-redundant
set $\mathbb N$ of $24,024$ crystal structures each has $>$800 atoms
(with protons) for a monomeric or $>$1000 atoms for a multimer, at
most 70\% sequence identity with any others, a resolution $\le$
3.5\AA~and an $R$-factor $\le 27.5$\%. The set $\mathbb N$ excludes
hyperthermophilic, membrane and nucleic acid binding proteins and the
size (number of atoms $n$) of its structures ranges from $833$ to
$171,552$ atoms. It is further divided into a set of monomerics
$\mathbb M$ with $8,974$ structures with sizes from $833$ to $44,200$
atoms and a set of multimers $\mathbb D$ with $15,050$ structures. Out
of $\mathbb M$ we select a subset $\mathbb M_{\mathrm f}$ of $1,766$
monomeric structures that have no gap in sequence, no compounds with
$>$5 atoms and $<$0.2\% missing atoms. The set $\mathbb M_{\mathrm f}$
is used to represent soluble proteins in free state and whose
structures have $1,004$ to $10,297$ atoms. Via the sequence
information in $\mathbb M_{\mathrm f}$ a set of extended and
energy-minimized conformations $\mathbb M_{\mathrm u}$ are generated
using CNS~\cite{cnsVersion1dot2} to study the changes in SES's
physical and geometrical properties upon protein unfolding. Protons
are added using the program {\sc reduce}~\cite{reduce1999} to any PDB
that lacks their coordinates. Please see the supplementary materials
for the preprocessing of PDBs for SES computation.

\subsection{The computation of solvent excluded surface}
Solvent excluded surface (SES) is composed of three types of areas:
solvent accessible area $a_{s}(i)$, torus accessible area $a_{t}(i,
j)$ and probe accessible area $a_{p}(i, j, k)$ where $a_{s}(i)$ is a
patch on the spherical surface of a single protein atom $i$, $a_{t}(i,
j)$ a toric patch defined by two atoms $i, j$ and $a_{p}(i, j, k)$ a
patch on the surface of a probe whose position is determined by three
atoms $i, j, k$. Geometrically $a_{s}$ is a convex area, $a_{p}$ a
concave one and $a_{t}$ a saddle area. Here we describe briefly the
key steps of our algorithm. It starts with the determination of all
the solvent accessible atoms ${\mathbb S}$ on both the exterior and
interior surfaces of a protein. For each pair of atoms in ${\mathbb
  S}$ and any third protein atom, we compute the probe defined by the
triple. Given the set of the computed probes ${\mathbb P}$, we
exhaustively search for the intersections between any pair of
probes. If there exists an intersection, the intersected area is
removed from further considerations. Given set ${\mathbb P}$ if any
two probes share a pair of atoms, we compute the torus defined by the
two probes and the two atoms. A subset of non-intersecting probes is
selected if there exist overlappings among them. Both $a_s$ and $a_p$
are computed by counting the number of exposed vertices on a spherical
surface that is represented by a set of uniformly-distributed $40,962$
vertices while $a_t$ is computed analytically. The algorithm is
implemented in C++ with pthread for parallel computation, SSE for
vector computation and Qt/openGL/GLSL for structure and surface
visualization. Please see the supplementary materials (Fig.~S1) for a
comparison of the surfaces by MSMS and our program. \\

In this study we set the probe radius to 1.4\AA~except for set
$\mathbb M$ over which SESs are computed twice using respectively
1.4\AA~and 1.2\AA. The SESs with 1.2\AA~radius are compared with those
with 1.4\AA~to see how the probe radius affects SES's physical and
geometrical properties. The atomic radii used are: C = 1.70\AA, N =
1.55\AA, O = 1.52\AA, S = 1.75\AA, H = 1.09\AA~and Se = 1.80\AA;

\subsection {The physical and geometrical properties of SES}
A variety of physical and geometrical properties defined on SES have
been computed to quantify their possible contributions to
protein-solvent interaction. Those we find to be relevant are listed
here and are called SES-defined properties for later reference.  Their
definitions rely on the assignment of charge and area to individual
atoms. 
\subsubsection{The physical properties of SES}
To each individual surface atom $i$ we assign an atomic SEA $a(i)$.
\begin{equation} \label{eq_area}
a(i)=a_{s}(i)+a_{t}(i)+a_{p}(i);\quad a_t(i)=\frac{\sum_{j} a_{t}(i,j)}{2},\quad a_p(i)=\frac{\sum_{j,k} a_{p}(i,j,k)}{3}.
\end{equation}
where $a_s(i)$, $a_t(i)$ and $a_p(i)$ are respectively the solvent
accessible, toric and probe areas for atom $i$.  For a protein we
define its SEA $A$, net surface charge $Q_s$ and average-partial
charge of the exposed atoms\footnote{{\em Solvent accessible, exposed}
  and {\em surface atoms} are used interchangeably in this paper.}
$\rho_s$.
\begin{align} 
A &= \sum_i a(i), \quad Q_s=\sum_i e(i), \quad i \in \mathbb S \nonumber \\
n_s &=|\mathbb S|, \quad \rho_s=\frac{Q_s}{n_s}
\end{align}
where $\mathbb S$ is the set of surface atoms with $n_s$ atoms, $e(i)$
the partial charge of atom $i$ whose value is taken from Charmm force
field~\cite{charmm2009}. The area-weighted surface charge $q_s$ and
area-weighted surface charge density $\sigma_s$ are defined as
follows.
\begin{equation}
q_s = \sum_i a(i) e(i), \quad \sigma_s=\frac{q_s}{A};\label{eq_charge_exposed}
\end{equation}
Set $\mathbb S$ could be further divided into two subsets: $\mathbb S
= \mathbb S^{+} \cup \mathbb S^{-}$ where $\mathbb S^{+}$ and $\mathbb
S^{-}$ are respectively the sets of atoms with $e(i) \ge 0$ and
$e(i)<0$. For both subsets we define their respective average-atomic
areas $\eta_s^+$ and $\eta_s^-$, area-weighted positive and negative
surface charges $q_s^+$ and $q_s^-$, and area-weighted surface charge
densities $\sigma_s^{+}$ and $\sigma_s^{-}$.
\begin{align} 
A^{+} &=\sum_i a(i), \quad n_s^+=|\mathbb S^+|, \quad \eta_s^{+}=\frac{A^+}{n_s^+}, \quad e(i) \ge 0 \nonumber \\
\quad q_s^{+} &=\sum_i a(i) e(i),\quad \sigma_s^{+}=\frac{q_s^+}{A^{+}}, \quad e(i) \ge 0 \nonumber \\
 \quad A^{-} &=\sum_i a(i),  \quad n_s^- =|\mathbb S^-|, \quad \eta_s^{-}=\frac{A^-}{n_s}, \quad e(i) < 0 \nonumber \\
\quad q_s^{-} &=\sum_i a(i) e(i), \quad \sigma_s^{-} =\frac{q_s^-}{A^{-}}, \quad e(i) < 0 \label{eq_pncharge}
\end{align}
where $n_s^+$ and $n_s^-$ are respectively the number of atoms in
$\mathbb S^{+}$ and $\mathbb S^{-}$. For the set of buried atoms
$\mathbb B$ in a protein we define its net charge $Q_b$ and
average-partial charge $\rho_b$.
\begin{align} 
Q_b&=\sum_j e(j), \quad j \in \mathbb B \nonumber \\
n_b &=|\mathbb B|,  \quad \rho_b =\frac{Q_b}{n_b} \label{eq_charge_buried}
\end{align}
where $n_b$ is the number of atoms in $\mathbb B$.  Please note that
the set of all the atoms for a protein $\mathbb A = \mathbb B \cup
\mathbb S$. The net charge $Q$, average-partial charge $\rho$ and
charge $Q_d$ of an electric dipole moment (or polarization vector)
$\vec {\mathbf P}$ \footnote{$\sigma_{pol} = \frac{Q_d}{V} = \vec
  {\mathbf P} \cdot {\mathbf n}$ where $V$ is the volume of the region
  enclosed by an SES, $\sigma_{pol}$ the charge density and ${\mathbf
    n}$ the surface normal.}  for a protein (dipole charge in short)
are defined as follows.
\begin{align} 
n&=n_b+n_s, \quad Q=Q_b+Q_s, \quad \rho =\frac{Q}{n} \nonumber \\ 
Q_d &= Q_s - \frac{Q}{2} \label {eq_charge_all}  
\end{align}
where $n = |\mathbb A|$ is the total number of atoms.\\

To evaluate the contribution of the hydrogen bonds between a protein
and its solvent to their interaction we divide $\mathbb S$ into two
subsets, $\mathbb S_{\mathrm o}$ of hydrophobic atoms and $ \mathbb
S_{\mathrm i}$ of hydrophilic atoms, with their respective surface
areas $A_o$ and $A_i$ defined as follows.
\begin{align}
\mathbb S &= \mathbb S_{\mathrm o} \cup \mathbb S_{\mathrm i} \nonumber \\
A_o&= \sum_i a(i), \quad i \in \mathbb O; \quad n_o=|\mathbb S_{\mathrm o}|, \quad \eta_o=\frac{A_o}{n_o} \nonumber \\ 
A_i&= \sum_j a(j), \quad j \in \mathbb I; \quad n_i=|\mathbb S_{\mathrm i}|, \quad \eta_i=\frac{A_i}{n_i} \label{eq_phiPho}
\end{align}
where $n_o$ and $n_i$ are their numbers of atoms, and $\eta_o$ and
$\eta_i$ their average-atomic areas. The protein atoms in $\mathbb
S_{\mathrm i}$ are either hydrogen bond donors or acceptors (H-bond
capable in short\footnote{In the rest of paper we use {\em H-bond
    capable atoms} and {\em hydrophilic atoms} interchangeably.})
while $\mathbb S_{\mathrm o}$ include the rest of atoms (see the
supplementary materials for their definitions). The area $A_i$ is
called the polar surface area of a protein in short.

\subsubsection{The geometry of SES}
The SEA $a(i)$ for atom $i$ is composed of three types of area:
$a_s(i)$, $a_t(i)$ and $a_p(i)$ with $a_s(i)$ a convex area defined by
a single atom $i$ while $a_p(i)$ a concave area defined by three atoms
$i, j, k$. For a set of surface atoms $\mathbb T$ we define a
convex-concave ratio $r_{cc}$ to estimate their overall surface
flatness and a sphere-volume over surface-volume $r_{pp}$ to measure
how tight a protein is packed.
\begin{align} \label{eq_ccRatio}
r_{cc}&=\frac{\sum_i a_p(i)}{\sum_i a_s(i)}, \quad i \in \mathbb T \nonumber \\
V_s&=\frac{4\pi}{3} (\frac{A}{4\pi})^{3/2}, \quad V_a=\frac{4\pi}{3}\sum_k r_k^3, \quad k \in \mathbb A \nonumber \\
r_{pp}&= \frac{V_s}{V_a}
\end{align} 
where $A$ is the total SEA, $V_s$ the surface-volume defined as the
volume of a sphere with the same surface area as $A$, and $V_a$ the
sum of atomic volumes over $\mathbb A$ and $r_k$ the radius of atom
$k$.  Since the concave area $a_p$ of a surface atom is determined by
triples of protein atoms, we could use $r_{cc}$ to estimate the
contribution to protein-solvent interaction of VDW attraction.

\section{Results and Discussion}
In this section we present the SES-defined physical and geometrical
properties over sets $\mathbb N, \mathbb M_{\mathrm f}$ and $\mathbb
M_{\mathrm u}$ and discuss their significance for protein-solvent
interaction.

\subsection{Surface charge and electric dipole and polar surface area}
Though it is well-documented that hydrophilic residues especially the
charged ones prefer to be on a protein surface, surface charge is
thought to be important for protein solubility\footnote{R.~P.~Feynman
  tried to explain the protein salt-out effect by assuming the
  existence of negative charges on protein surfaces.``The molecule
  (protein) has various charges on it, and it sometimes happens that
  there is a net charge, say negative, which is distributed along the
  chain'', The Feynman Lectures on Physics, page 7--10, Vol.2.} and
most important physical and chemical properties of a protein are
ultimately related to the electrostatic interactions among its
composing atoms and with other molecules such as
solvent~\cite{Perutz1187, Warshel20061647}, to our knowledge no
large-scale surveys of surface charges have been reported. With atomic
SEA and the separation of surface atoms from buried ones and the
division of atoms into different subsets according to their
physical-chemical properties (Eqs.~1--7), it is possible to evaluate
the contributions of surface area and charge to protein-solvent
interaction by performing a statistical analysis over known structures
on physical and geometrical properties defined by them. In theory,
evolution must have optimized soluble proteins for best interacting
with water and since the latter is a protic solvent that prefers
anions over cations we expect that folding into a native state will
turn a soluble protein into an anion with positive charges buried
inside. Indeed we find that all the $24,024$ structures in $\mathbb N$
have negative net surface charges (negative $Q_s, q_s$ and $\rho_s$)
and positive net buried charges (positive $Q_b$ and
$\rho_b$)\footnote{Using the criteria listed in section 2.1 two
  structures (PDBIDs:3odv and 4uj0) have positive surface
  charges. Both of them are membrane-permeable toxins composed of
  several peptides. Another membrane-permeable small monomeric protein
  (PDBID:1bhp) with 658 atoms including protons also has positive
  surface charge.  One structure (PDBID:1w3m, an antibiotic named
  tsushimycin) has very small net negative buried charge, $Q_b= -1.68$
  coulomb.  They are excluded in the present study for easy
  exposition.}  (Fig.~1a) and more strikingly the difference in mean
between average-partial charges $\rho_e$ and $\rho$ is more than
17-fold. It is equivalent to a 17-fold increase in negativity on
average when a protein folds into a native state.  In addition the net
surface charges for the structures in $\mathbb M$ remain to be
negative even when the probe radius is reduced from 1.4\AA~to
1.2\AA~(see supplementary materials).  More hydrophobic atoms become
exposed with a smaller probe radius. Furthermore negative surface
charges increase with protein size\footnote{In this paper protein size
  could mean either $n$ or $n_s$ or $A$ since they are related to each
  other via simple relationships (supplementary materials).} via
well-fitted power laws and their enrichment is apparent upon
folding. As shown in Table 1, the difference in $\rho_s$s between the
folded structures ($\mathbb M_{\mathrm f}$) and unfolded ones
($\mathbb M_{\mathrm u}$) is $>$30-fold on average. The extended
conformations in $\mathbb M_{\mathrm u}$ may deviate from the real
unfolded states existent in a typical experimental setting and thus
the geometrical and electrical properties computed on them may have
errors. However the large difference in $\rho_s$ between $\mathbb
M_{\mathrm f}$ and $\mathbb M_{\mathrm u}$ supports the relevance of
surface charge to protein-solvent interaction.  The findings described
above together show that surface charge is a universal property
importance for protein-solvent interaction.\\

\begin{figure}[htb]
\centering
\hspace{-0.9cm}\subfloat[Average-partial charges $\rho_s, \rho_b$ and
  $\rho$]{\label{fig:surfaceCharge}
  \includegraphics[width=0.5\textwidth,
    height=0.3\textwidth]{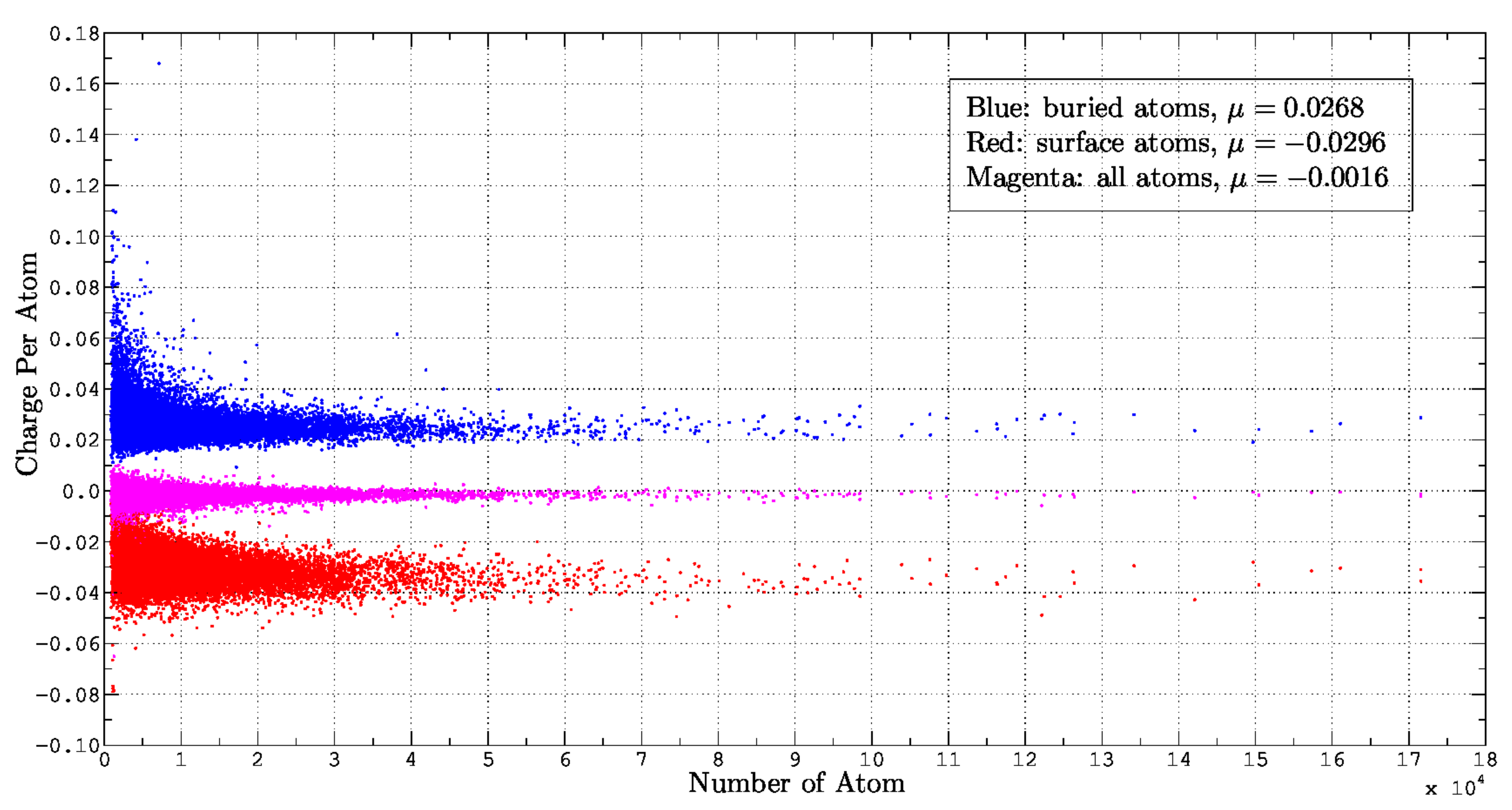}}
\subfloat[Dipole charge $Q_d$]{\label{fig:minVdwSurf}
  \includegraphics[width=0.5\textwidth,
    height=0.295\textwidth]{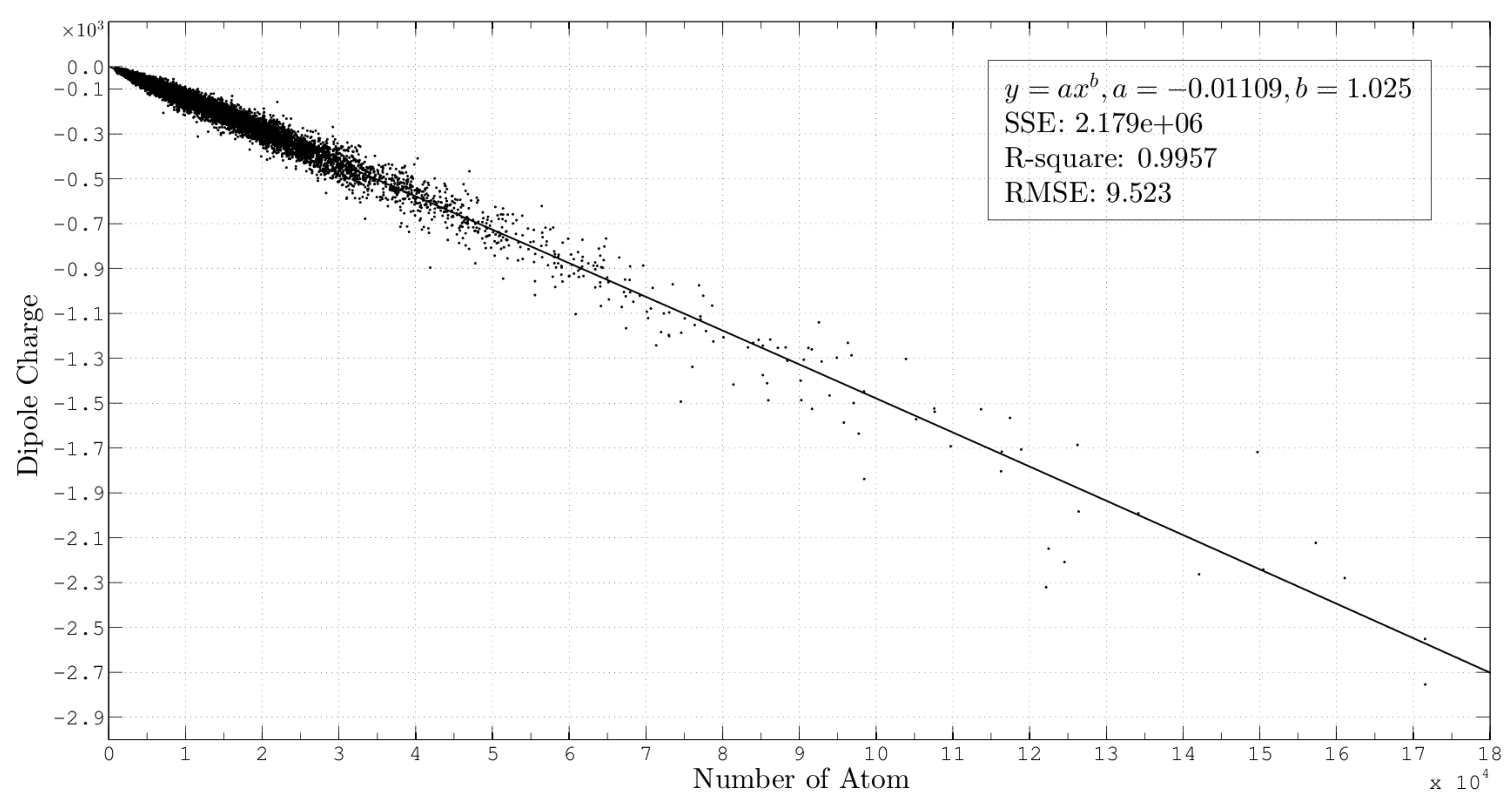}}
\caption { \small{ \textbf {Average-partial charges $\rho_s, \rho_b,
      \rho$ and dipole charge $Q_d$}. {\small {\bf(a)} The
      average-partial charges $\rho_s$ (colored in red), $\rho_b$
      (blue) and $\rho$ (magenta) over $\mathbb N$. The mean value
      $\mu$s for $\rho_b, \rho, \rho_s$ are respectively $0.0268,
      -0.0016$ and $-0.0296$, and the difference between the $\mu$s
      for $\rho_s$ and $\rho$ is $>$17-fold.  The y-axis is
      average-atomic charge in coulomb per atom. {\bf (b)} Dipole
      charge $Q_d$s over $\mathbb N$.  The change of $Q_d$s with
      protein size ($n$) could be fitted very well either to a power
      law $Q_d=an^b$ with a ${\mathrm R_{square}} =0.9957$ and
      $b=1.025$ or a linear equation (data not shown). The y-axis is
      dipole charge in coulomb. The x-axes in both ({\bf a, b}) are
      $n$, the total number of atoms in each structure. } }}
\label{fig:surfaceCharge} 
\end{figure} 

\begin{table}[t]
\centering
\small{
\begin{tabular}{|c|c|c|c|c|c|}
\hline
 Structures & Exposed ($\rho_s$) & Buried ($\rho_b$) & Total ($\rho$) & Exposed ($\rho_{_\mathrm {adj}, s}$) & Buried ($\rho_{_\mathrm {adj}, b}$)\\ \hline
$\mathbb M_{\mathrm f}$   & -26.6, 6.1  &  27.6,  7.9  & -1.3,   2.4 & -25.3,  4.7 & 28.9, 8.3 \\\hline
$\mathbb M_{\mathrm u}$ & 0.7402, 2.9 & -13.8, 33.4  & 0.2508, 2.5 & 0.4894, 1.2 &-14.1, 33.7 \\\hline
\end{tabular}
}
\caption{{\small\label{tbl:results} \textbf {The average-partial
      charges over the sets of exposed ($\mathbb S$), buried ($\mathbb
      B$) and total atoms ($\mathbb A$) for the folded ($\mathbb
      M_{\mathrm f}$) and unfolded ($\mathbb M_{\mathrm u}$)
      structures}. The two numbers in each cell are respectively mean
    and standard deviation. The unit is $10^{-3} \times $ coulomb per
    atom. Some structures in $\mathbb M_{\mathrm f}$ may have no
    coordinates for the free amine groups at their N-termini and that
    leads to a negative mean for the $\rho$s for the folded
    structures.  The adjusted average-partial charges $\rho_{_\mathrm
      {adj}, s}$ and $\rho_{_\mathrm {adj}, b}$ are computed using the
    adjusted surface charge $Q_d = Q_S - \frac{Q} {2}$ for the folded
    and unfolded structures. }}
\end{table}
All the structures in $\mathbb N$ have small nonzero net charges
($Q$s) with an average of $\rho=-0.0016$ coulomb per atom
(Fig.~1a). The small negative $\rho$s are due in part to the lack of
coordinates for the free amine groups at protein's N-termini. To make
a structure neutral in charge we define an adjusted surface charge
$Q_d = Q_s - \frac{Q} {2}$. It is the charge for a permanent electric
dipole moment $\vec {\mathbf P}$. As shown in Fig.~1b, $Q_d$ decreases
almost linearly with protein size $n$, and most interestingly $\vec
{\mathbf P}$ changes its direction upon protein folding (Table 1),
that is, the signs of the surface charges $Q_s$s and $Q_d$s of the
unfolded conformations change from being positive to negative upon
folding. Thus $\vec {\mathbf P}$ is likely to be a universal quantity
important for protein-solvent interaction. Taken together, it is clear
that folding into a native state in solution turns a protein into a
capacitor with a net negative charge on its SES, the outer surface of
a capacitor, to maximize its attraction to the
solvent~\cite{KADill2008}. Except for the charged side chain atoms
Charmm partial charges~\cite{charmm2009} for a residue have zero net
charge for its subgroups of bonded atoms and thus except for the
regions where the charged side chain atoms are located, there must
exist a 2D manifold (the inner surface of the capacitor) inside a
soluble protein that encloses a set of atoms with zero net charge. A
model of alternative layers of negative and positive charges has been
alluded previously in molecular dynamic (MD)
simulation~\cite{Simonson1995}. Though the details of such a
multi-layer model could not be worked out without knowing the 2D
manifold for each layer, if proved to be correct it may provide an
explanation to why the protein interior is like a medium of high
polarizability. \\

The strength of the interaction between a polar solute and a protic
solvent such as water is determined also by the hydrogen bonds between
them while the strength of the latter depends on both distance and
direction. Thus the larger the SEA of a hydrophilic atom has the
stronger of its interaction with water since a larger SEA is less
disruptive to water structure and thus causes less loss of solvent
entropy.  In theory the optimization via evolution must have maximized
such polar surface areas. Indeed as shown in Fig.~2a the
average-atomic areas ($\eta_i$s) over the sets ($\mathbb S_{\mathrm
  i}$s) of hydrophilic atoms over $\mathbb N$ are on average 1.75-fold
larger than those for the corresponding $\eta_o$s for hydrophobic
atoms. In fact not only the hydrophilic atoms have larger surface
areas the $\eta_s^{-}$s for the sets ($\mathbb S^{-}$s) of the surface
atoms with negative partial charges are on average larger than the
$\eta_s^+$s (Fig.~2a and Table 2), this is consistent with the
preference of water as a protic solvent for anions over
cations. Moreover though both $A_o > A_i$ and $A^{+} > A^{-}$ for
either folded or unfolded structures, their ratios decrease upon
folding (Table 2). Taken together it suggests that folding into a
native state not only turns a protein into a capacitor but also
maximizes its hydrogen bonding interaction with solvent with as little
disruption to solvent structure as possible.\\

Though both average-partial charge and average-atomic area contribute
to protein-solvent interaction their changes with protein size are
somewhat complicated. For example, both of them decrease with protein
size by power laws and their decreases are faster for small structures
with $n < 2\times10^4$. In addition the distributions around their
means are not symmetrical especially for average-atomic areas
(Figs.~1a and 2a). The non-uniformity in their decreases with protein
size implies that neither of them alone could properly describe
protein-solvent interaction because its strength per exposed atom is
expected to be statistically independent of size. In contrast to
average-partial charge and average-atomic area, area-weighted surface
charges ($q^{+}_s$, $q^{-}_s$ and $q_s$) change linearly with protein
size (supplementary materials) and moreover area-weighted surface
densities ($\sigma_s^+, \sigma_s^-$ and $\sigma_s$) are almost
independent of size (Fig.~2b). In addition the distributions around
their means are rather symmetrical as indicated by the very small
differences between their means and medians even for small-sized
structures in sets $\mathbb M_{\mathrm f}$ and $\mathbb M_{\mathrm u}$
(Table 3).  Taken together it shows that SES properties that are
functions of either charge or both surface area and charge provide a
better description to protein-solvent interaction than area alone
does. Previous applications~\cite{CHOTHIA19761, Sharp106,
  Ashbaugh1999, charmm2009} have focused mainly on area alone (either
ASA or SEA) and their usefulness for quantifying protein-solvent
interaction remains to be controversial~\cite{Tanford1979,
  Zhou2013,Pettitt2014}. As shown here, our large-scale analyses of
SEAs and SES-defined physical properties over both folded and unfolded
structures demonstrate the importance of surface charge in
protein-solvent interaction and also explains the inadequacy of using
area alone for its evaluation. Furthermore, the statistical values for
SES-defined properties and power laws obtained for the set of
monomerics ($\mathbb M$) could be used to quantify the changes induced
by ligand binding where the ligand is either DNA or small-molecule
compound or another protein~\cite{sesProtLigand}.
   
\begin{figure}[htb]
  \centering
  \hspace{-0.9cm}\subfloat[Average-atomic
    area]{\label{fig:surfaceCharge}
    \includegraphics[width=0.5\textwidth,height=0.295\textwidth]{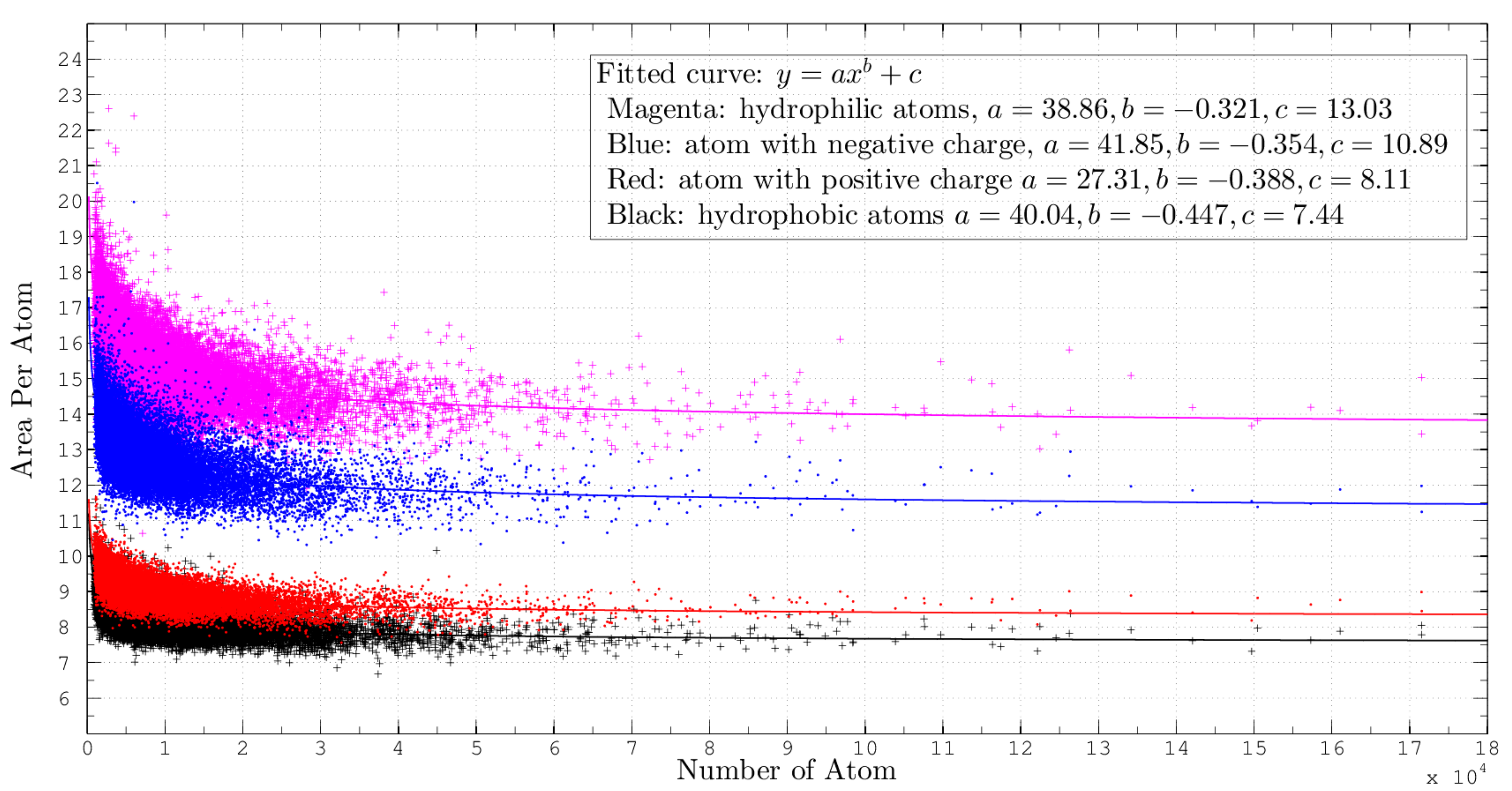}}
  \subfloat[Area-weighted surface charge
    density]{\label{fig:minVdwSurf}
    \includegraphics[width=0.5\textwidth,
      height=0.3\textwidth]{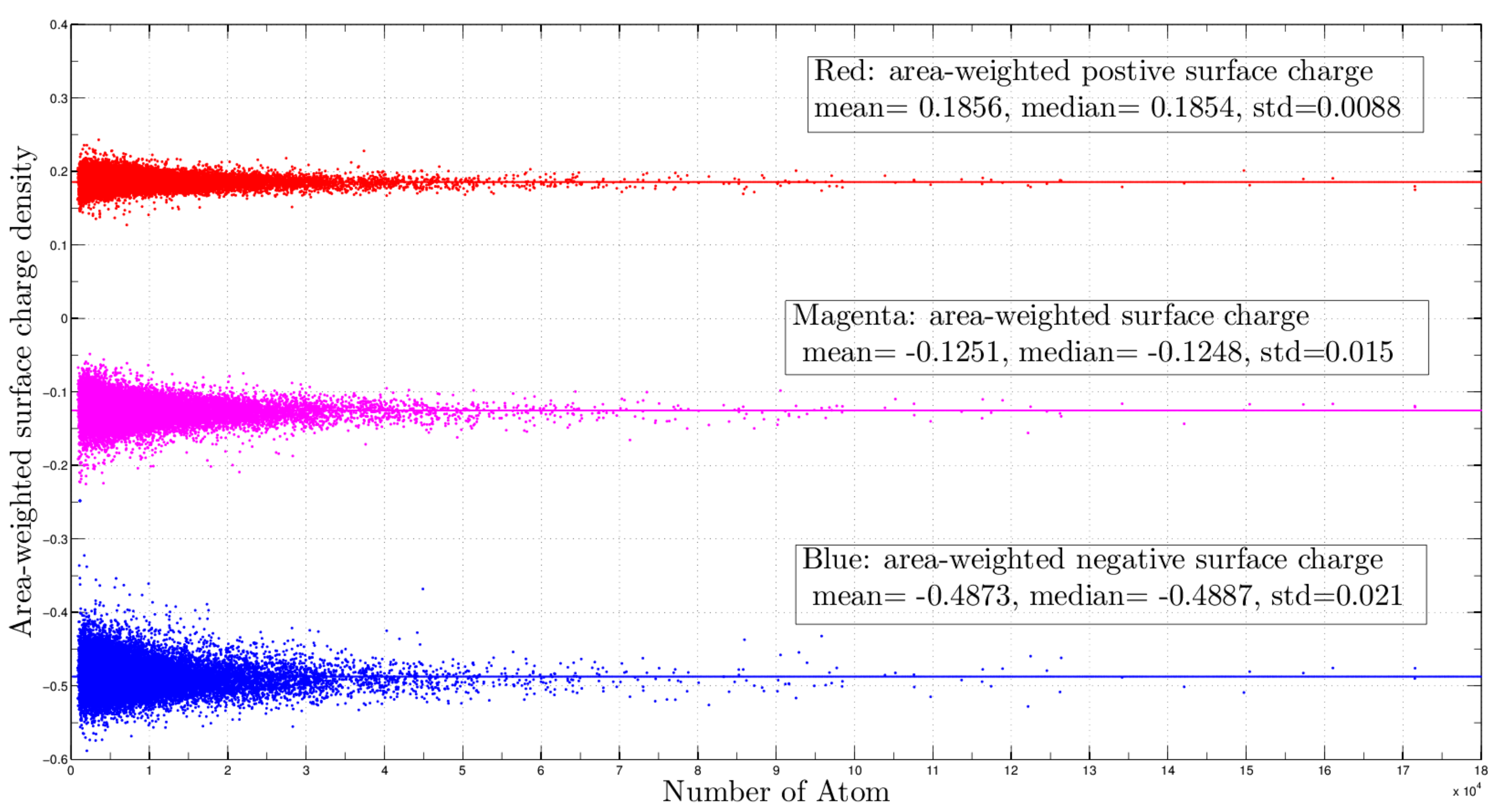}}
  \caption{ \textbf {Average-atomic areas and area-weighted surface
      charge densities.} {\small {\bf(a)} The average-atomic areas
      $\eta_i, \eta_o$ and $\eta_s^+, \eta_s^-$ over $\mathbb N $. The
      fitted power laws, $y=ax^{b} + c$, shown in the insert all have
      negative $b$s and thus the values of parameters $c$ could be
      used to compare their average-atomic areas. For example, the
      parameter $c$ for $\eta_i$s is 1.75-fold larger than that for
      $\eta_o$s. The y-axis is average-atomic area with a unit of
      \AA$^2$ per atom. {\bf(b)} The area-weighted surface charge
      densities of $\sigma_s^+, \sigma_s^-$ and $\sigma_s$. The three
      lines indicate their respective means.  The y-axis is
      area-weighted surface charge density in coulomb per atom. The
      x-axes in ({\bf a, b}) are the total number of atoms in a
      protein.}  }  \label{fig:surfaceCharge}
\end{figure} 

\begin{table}[t]
\centering
\small{
\begin{tabular}{|c|c|c|c|c|c|c|}
\hline
 Structures & $\eta_i$  & $\eta_o$  & $\frac{A_o}{A_i}$ &  $\eta^{+}$  & $\eta^{-}$ & $\frac{A^{+}}{A^{-}}$ \\ \hline
$\mathbb M_{\mathrm f}$   & 19.1792     & 12.3654 & 1.2102 & 12.9688 & 17.061 & 1.1918 \\\hline
$\mathbb M_{\mathrm u}$  &16.1297     &  8.5199 & 1.5714 & 9.3988 &13.2806  & 1.4865 \\\hline
\end{tabular}
}
\caption{{\small\label{tbl:results} \textbf {The average-atomic areas
      of the folded ($\mathbb M_{\mathrm f}$) vs unfolded ($\mathbb
      M_{\mathrm u}$) structures}. In addition to the average-atomic
    areas the ratios $\frac{A_o}{A_i}$ and $\frac{A^{+}}{A^{-}}$ are
    listed respectively in the third and last columns. The unit for
    average-atomic areas is \AA$^2$ per atom.}}
\end{table}

\subsection{The geometry of SES}
One advantage of SEA over ASA is that the former includes both convex
and concave areas while the latter has only convex ones. Using SES we
could define a geometrical property $r_{cc}(i) = \frac{a_p(i)}{
  a_s(i)}$ for each surface atom $i$ to estimate its local flatness
and $r_{cc} = \frac{\sum_i a_p(i)}{\sum_i a_s(i)}, i \in \mathbb T$
(Eq.~8) for a set of atoms $\mathbb T$ to estimate the overall
flatness of their total surface. The ratio $r_{cc}$ is possibly
related to the VDW interaction~\cite{Pettitt2014, Baldwin2014} between
surface atoms and solvent since we have $d_{ap}(i) = r_a(i) + r_p$
where $d_{ap}(i)$ is the inter-atom distance between protein atom $i$
and a solvent atom, $r_a(i)$ the VDW radius for atom $i$ and $r_p$
probe radius.  In addition the larger the ratio is, the more flat the
surface, the smaller the surface tension and thus stronger the
interaction with solvent. We also expect that the natural selection
must have optimized protein-solvent interaction in terms of surface
geometry via the maximization of the concave-convex ratio for
hydrophobic atoms since VDW attraction is assumed to be the main
factor for their solvation in polar solvent~\cite{Baldwin2014}. Indeed
as shown in Fig.~3a the $r_{cc}$s for the sets of the hydrophobic
atoms ($A_o$s) over $\mathbb N$ are about 40\% larger than those for
the $A_i$s. Moreover $r_{cc}$s increase with protein size via
well-fitted power laws (Fig.~3a) and the increase becomes more slowly
when the number of atoms $n > 2.0 \times 10^4$. With more and more
surface atoms it becomes increasingly possible to form flatter and
flatter surface and consequently better and better protein-solvent
interaction as far as surface geometry is concerned. Most
interestingly, in stark contrast with the $r_{cc}$s over $\mathbb
M_{\mathrm f}$, the same $r_{cc}$s over $\mathbb M_{\mathrm u}$ are
several-fold smaller and do not change with protein size
(Fig.~3b). Taken together it shows that concave-convex ratio $r_{cc}$
is a geometrical property relevant to protein-solvent interaction
possibly via the VDW attraction between surface atoms and solvent.

\begin{figure}[htb]
  \centering
  \hspace{-0.9cm}\subfloat[The concave-convex ratios over $\mathbb N$
  ]{\label{fig:sesGeometry}
    \includegraphics[width=0.5\textwidth,height=0.3\textwidth]{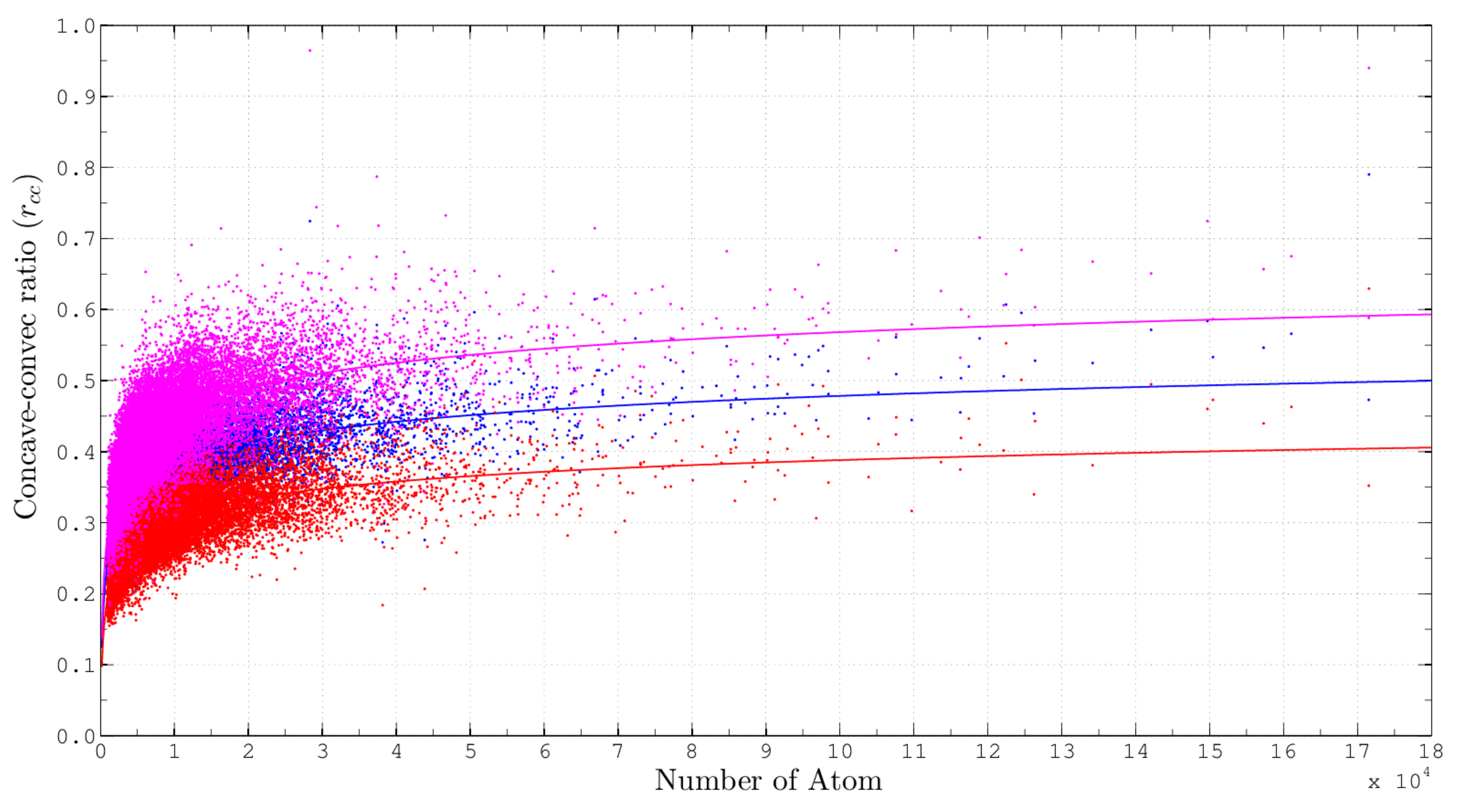}}
  \subfloat[The concave-convex ratios over $\mathbb
    M_{\mathrm f}$ and $\mathbb M_{\mathrm u}$
  ]{\label{fig:minVdwSurf} \includegraphics[width=0.5\textwidth,
      height=0.3\textwidth]{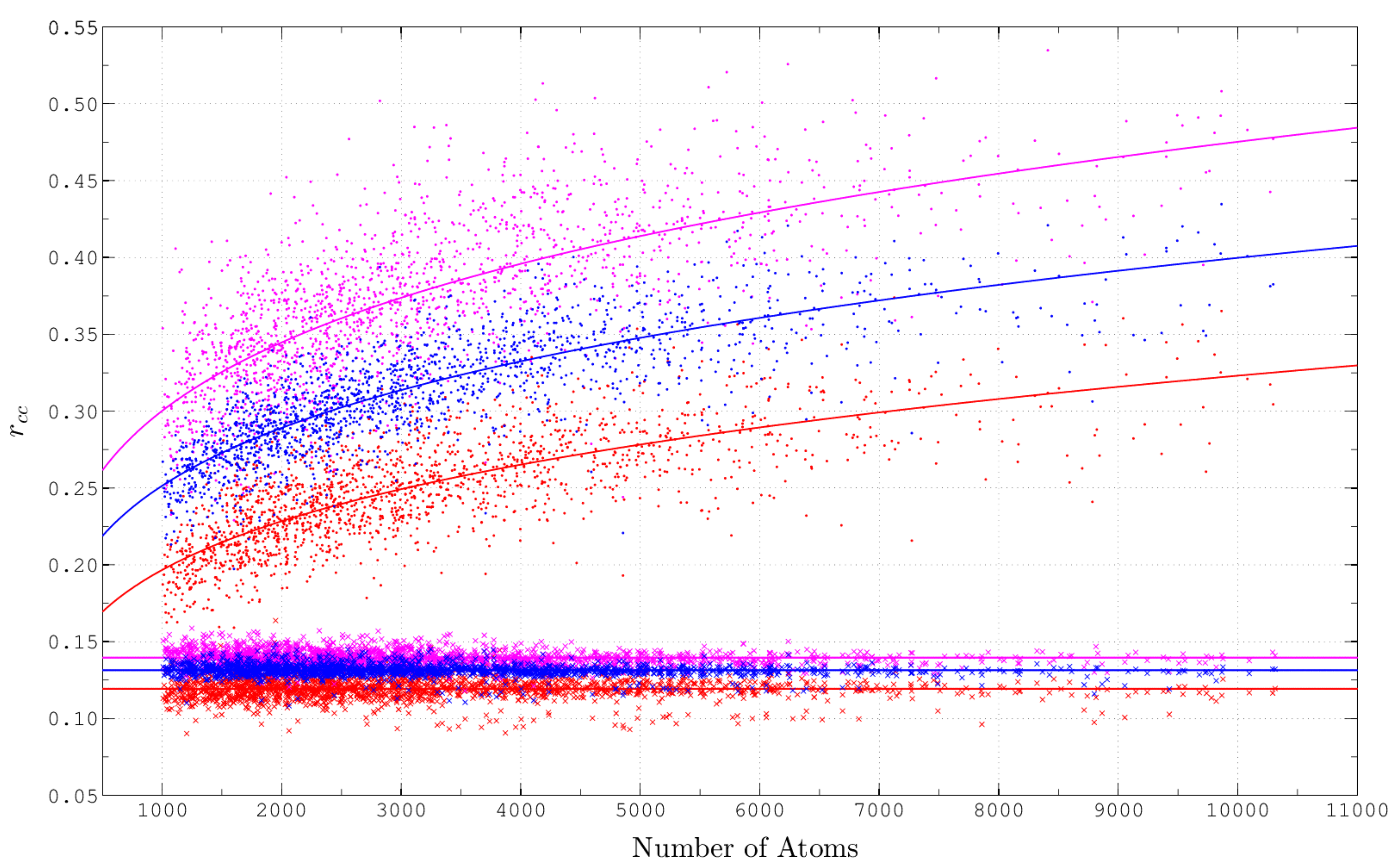}}
  \caption{\textbf {Concave-convex ratio $r_{cc}$}. {\small {\bf(a)}
      The $r_{cc}$s for the hydrophilic (set $\mathbb S_{\mathrm i}$
      and colored in red), hydrophobic ($\mathbb S_{\mathrm o}$ in
      blue ) and all the atoms (in magenta) over $\mathbb N$. The
      three curves represent the fitted power laws $r_{cc}(n)=a n^b +
      c$ shown in the insert. The $b$s are all negative so the
      parameter $c$s where $c = \lim_{n \rightarrow \infty}r_{cc}(n)$
      could be used to compare their concave-convex ratios. {\bf (b)}
      The $r_{\mathrm cc}$s of both folded and unfolded
      structures. The three curves are fitted power laws respectively
      for the sets of hydrophilic (colored in red), hydrophobic (in
      magenta) and all the atoms (in blue) over $\mathbb M_{\mathrm
        f}$ while the three lines indicate their means over $\mathbb
      M_{\mathrm u}$. } }
\end{figure} 

\begin{table}[t]
\centering
\small{
\begin{tabular}{|c|c|c|c|c|}
\hline
 Structures & $\sigma_s^{+}$ &  $\sigma_s^{-}$  &  $\sigma_s$ &  $\sigma_s^{+} - \sigma_s^{-}$  \\ \hline
$\mathbb M_{\mathrm f}$ & 10.13, 10.11, 0.60  & -22.42, -22.39, 1.46 &-12.29, -12.23, 1.67 & 32.54, 32.55, 1.48 \\ \hline
$\mathbb M_{\mathrm u}$ & 10.50, 10.47, 0.43 & -18.29, -18.24, 0.84 & -7.79, -7.75, 0.92 & 28.80, 28.74, 0.96 \\ \hline
\end{tabular}
}
\caption{{\small\label{tbl:results} \textbf {The area-weighted surface
      charge and area-weighted surface dipole densities of the folded
      vs unfolded structures}. The three numbers in each cell are
    respectively mean, median and standard deviation with a unit of
    $10^{-2} \times$ coulomb. If we assume that the structures in
    $\mathbb M_{\mathrm u}$ are the representatives of unfolded
    states, then folding into a native state makes the surface
    slightly more positive but drastically more negative.}}
\end{table}

\subsection{Hydrophobicity scale}
As described above the changes with protein size or upon unfolding of
most SES-defined physical and geometrical properties follow
well-fitted power laws. These properties are defined with the
assignment of area and charge to individual atoms rather than
individual residues. However the pertinent experimental data such as
hydrophobicity scales are available only at residue-level. In this
section we first describe the correlations between five known
hydrophobicity scales~\cite{Janin1979, KYTE1982105,EWscale1984,
  GES1986} and a dozen of physical and geometrical properties computed
for each of 20 amino acid residues (Table 4) and then discuss their
significance for protein-solvent interaction in general and folding in
particular since hydrophobic effect is thought to be the driving force
for the latter~\cite{Kauzmann1987, Dill1990}.  We use Kyte-Doolittle
scale as a reference since it incorporates both experimental data and
computational surveys from several sources. The seven SES-defined
electrical properties include net positive and negative surface
charges, $Q^{+}$ and $Q^{-}$, area-weighted positive and negative
surface charges, $q^{+}$ and $q^{-}$, as well as their differences
$Q^{+}-Q^{-}$ and $q^{+}-q^{-}$ and the area-weighted surface charges
($q_i$s) of the hydrophilic atoms for a residue. The five geometrical
properties include total SEA, the SEAs of hydrophilic, positive and
negative-charged atoms as well as concave-convex ratio $r_{cc}$. Their
correlations with the five scales are assessed by the goodness of
fitting to either a linear equation or a power law. As shown in Table
4 the seven electrical properties in general fit to the five scales
better than the five geometrical properties such as $A$ and $r_{cc}$
do.  The best correlations exist between four SES-defined electric
properties, $Q^{-}, q^{+}, (q^{+}- q^{-})$ and $Q^{+}-Q^{-}$, and the
five scales. By comparison SEA ($A$) alone does not fit as well to the
five scales. In fact except for $r_{cc}$ the fitting between SEA and
the five scales are the worst among the twelve properties. Previously
the contribution of surface to protein-solvent interaction has been
evaluated using mainly areas including both surface area and buried
average area~\cite{Rose834}. Furthermore it is assumed that there is a
hydrophobic component $\Delta G_{\mathrm {hyd}}$ of $\Delta
G_{\mathrm{solv}}$ that increases linearly with area $A$, and the
assumption is widely used in implicit solvent
models~\cite{Pettitt2014}.  However the results presented here and in
section 3.1 show that surface charge and electrical properties are
more important than surface area in terms of their contributions to
protein-solvent interaction. On the other hand, since strong
correlations exist between either $q^{+}$ or $(q^{+}- q^{-})$ and the
five scales we could interpret the fitted solvation parameters
($\sigma_i$s)~\cite{Eisenberg1986, OOI1987} in solvation free
energy-surface area relation, $\Delta G_\mathrm{solv} = \sum_{i}
\sigma_{i} \ ASA_{i}$ as surface charges since $ASA_{i}$ is to a good
extent proportional to the SEA for the same residue.

\begin{figure}[htb]
\centering \subfloat[Kyte-Doolittle vs
  $q_s^{+}$]{\label{fig:kdVsPaaOverSaa}
\hspace{-0.6cm}
  \includegraphics[width=0.5\textwidth, height=0.25\textwidth]{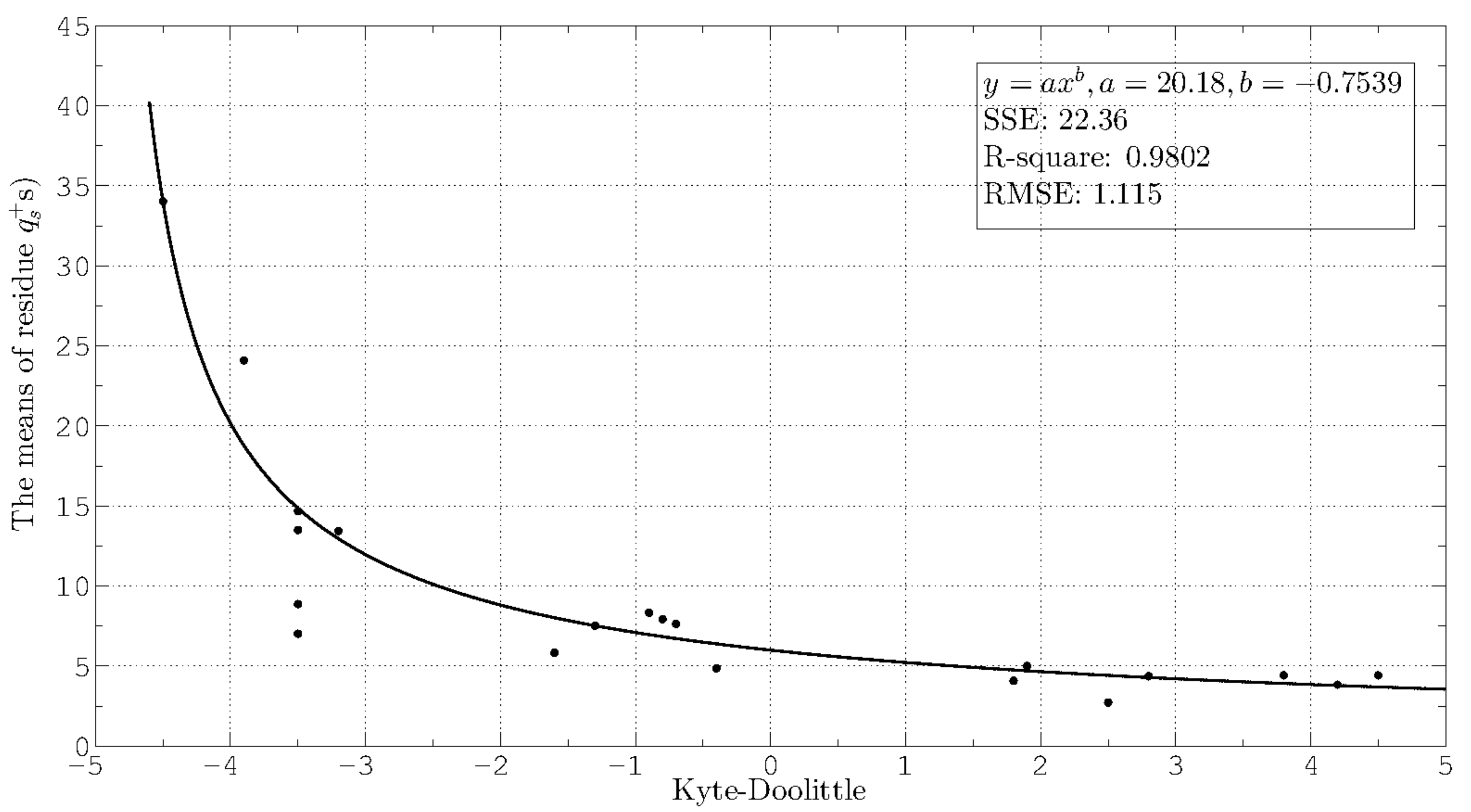}}
 \subfloat[Kyte-Doolittle vs ratio $r_{cc}$]{\label{fig:kdVsDipole}
\includegraphics[width=0.5\textwidth, height=0.253\textwidth]{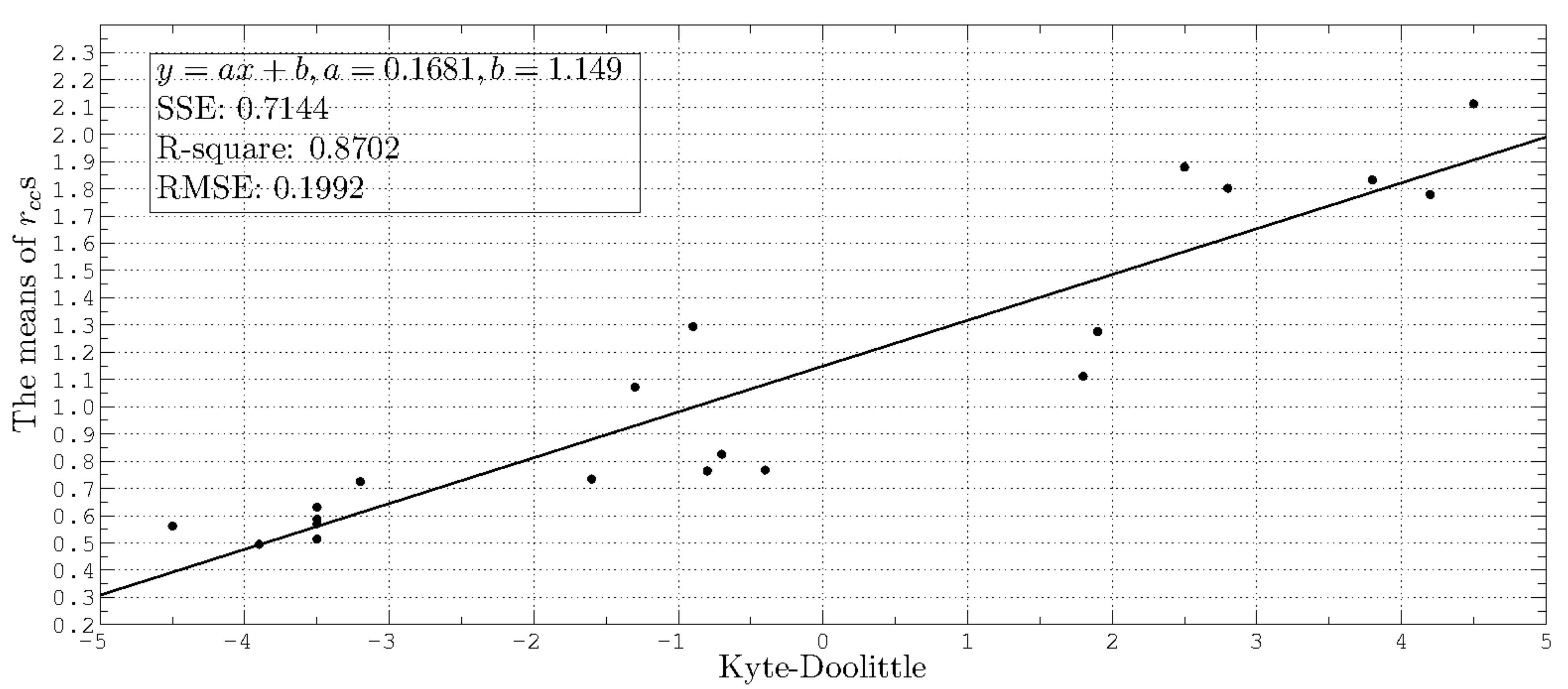}}
\caption{ \textbf {The means of net positive surface charges and
    concave-convex ratios of 20 residues versus Kyte-Doolittle
    hydrophobicity scale. } {\small {\bf(a)} The means of net positive
    surface charges ($q_s^{+}$s) of 20 residues versus Kyte-Doolittle
    scale. The curve is a best-fitted power law. The y-axis is the
    means of $q_s^{+}$s for 20 residues in coulomb. {\bf (b)} The
    means of concave-convex ratios ($r_{cc}$s) of 20 residues versus
    Kyte-Doolittle scale. The line is a best-fitted linear
    equation. The y-axis is the means of $r_{cc}$s for 20
    residues. The x-axes in both {\bf(a, b)} are Kyte-Doolittle
    scale.}}
\label{fig:surfaceCharge} 
\end{figure} 

\begin{table}[t]
\centering
\small{
\begin{tabular}{|c|c|c|c|c|c|c|c|c|c|c|c|c|}
\hline
  &    $Q^{+}$  & $Q^{-}$  & $A$      & $A^{+}$  & $A^{-}$  &$A_{i}$  & $q_s^{+}$  & $q_s^{-}$   &$q_{i}$   & $p_r$ & $p_s$ & $r_{cc}$\\ \hline
KD&   0.92$^b$ & 0.96$^b$ & 0.94$^b$ & 0.93$^b$ & 0.94$^b$ & 0.89$^b$ & 0.98$^b$ & 0.90$^b$ & 0.85$^b$ & 0.94$^b$ & 0.96$^b$ & 0.87$^a$ \\\hline
EW&   0.86$^a$ & 0.92$^b$ & 0.68$^a$ & 0.91$^b$ & 0.52$^b$ & 0.85$^a$ & 0.96$^b$ & 0.83$^a$ & 0.82$^a$ & 0.85$^a$ & 0.91$^b$ & 0.71$^a$ \\\hline
GES&  0.83$^a$ & 0.89$^a$ & 0.71$^a$ & 0.90$^b$ & 0.52$^a$ & 0.88$^a$ & 0.85$^a$ & 0.93$^a$ & 0.82$^a$ & 0.89$^a$ & 0.89$^a$ & 0.51$^a$ \\\hline
JANIN&0.95$^b$ & 0.91$^b$ & 0.68$^b$ & 0.79$^b$ & 0.88$^b$ & 0.94$^b$ & 0.96$^b$ & 0.87$^b$ & 0.84$^b$ & 0.94$^b$ & 0.92$^b$ & 0.90$^a$ \\\hline
EXP&  0.85$^a$ & 0.93$^b$ & 0.65$^a$ & 0.65$^a$ & 0.93$^b$ & 0.65$^a$ & 0.85$^b$ & 0.93$^b$ & 0.93$^b$ & 0.90$^b$ & 0.93$^b$ & 0.63$^a$ \\\hline
Average&0.88   & 0.92     & 0.73    & 0.84     & 0.76     & 0.84     & 0.92     & 0.89    & 0.85      & 0.90     & 0.92    & 0.72 \\\hline
\end{tabular}
}
\caption{{\small\label{tbl:results} \textbf {The five hydrophobicity
      scales versus twelve SES-defined physical and geometrical
      properties}. The five scales are respectively Kyte-Doolittle
    (KD)~\cite{KYTE1982105}, Eisenberg-Weiss(EW)~\cite{EWscale1984},
    Goldman-Engelman-Steitz(GES)~\cite{GES1986},
    Janin(JANIN)~\cite{Janin1979} and experimental hydrophobicity
    scales. The data for experimental scale is taken from Table xx of
    Kyte-Doolittle paper~\cite{KYTE1982105}. The twelve properties are
    computed over sets of atoms belonging to individual
    residues. Among them the seven properties, $A, A^{+}, A^{-},
    A_{i}, q^{+}, q^{-}, q_{i}$ and $r_{cc}$ (Eqs.~2--8), are defined
    in section 2. The first two columns are respectively positive
    surface charge $Q^{+}=\sum_i e(i), e(i) > 0$ and negative surface
    charge $Q^{-}=\sum_i e(i), e(i) < 0$.  The 9th column $q_i$ is the
    total area-weighted surface charge for all the exposed hydrophilic
    atoms of a residue. The 10th and 11th columns are respectively
    $p_r =Q^{+} - Q^{-} = \sum_i e^+(i) - \sum_j e^-(j)$ and $p_s
    =q^{+} - q^{-} = \sum_i a(i) e^+(i) - \sum_j a(j) e^-(j)$ where
    $i$ is an exposed atom of a residue with positive partial charge
    and $j$ an exposed atom with negative charge. For each residue we
    first compute the twelve properties over set $\mathbb M_{\mathrm
      f}$ and then compute their means. Finally the correlations
    between their means and the five scales are assessed by fitting
    them to either a linear equation $y=ax+b$ (indicated by
    superscript $^a$) or a power law $y=ax^b$ (indicated by
    superscript $^b$) where $x$ is a hydrophobicity scale and $y$ one
    of the twelve properties.  The number in each cell is the
    coefficient of determination ${\mathrm R_\mathrm {square}}$.  The
    last row is the average of ${\mathrm R_\mathrm {square}}$s.}}
\end{table}

\subsection{The statistical distributions and power laws for SES-defined physical and geometrical properties}
At present the details of protein-solvent interaction could only be
obtained through all-atom MD simulation with either explicit or
implicit solvent models. However MD with explicit solvent suffers from
convergence problem while implicit models rely on a prior values for
dielectric constants especially the dielectric constants near a
protein surface or inside a protein~\cite{Warshel20061647}.  For
example accurate dielectric constant for protein surface is the key
for the computation of solvation free energy~\cite{Warshel20061647}
via electrostatic interaction.  However the accurate determination of
such dielectric constants remains to be a very challenging problem at
present. As described above we have found a dozen of SES-defined
electrical and geometrical properties that are likely to be important
for protein-solvent interaction. Their statistical distributions and
the power laws governing their changes with protein size or upon
protein unfolding obtained on large sets of known structures should be
useful for the quantification of solvation and hydrophobic effect
using well-known theories on anion solutes in protic
solvent~\cite{Onsager1936, CKmodel1957} or PLDL solvent
model~\cite{WARSHEL1976227, Warshel20061647}. In addition, the
statistical values and power laws could be used to restraint the
folding space of a protein and thus could serve as a term in an
empirical scoring function for protein structure
prediction~\cite{PROT:PROT21715} or a quantity for quality control in
structure determination~\cite{Kota15082011}.

\section{Conclusion}
A robust and accurate algorithm for the computation of solvent
excluded surface (SES) has been developed and applied to a large set
of soluble proteins with crystal structures. We discover that all the
soluble proteins have net negative surface charge and thus soluble
proteins behave like a capacitor in solution. We have also identified
a dozen of SES-defined physical and geometrical properties that are
relevant to protein-solvent interaction based on their changes with
protein size and upon protein unfolding as well as the strong
correlation between them and five known hydrophobicity scales on a
residue level. Most interestingly in contrast to previous emphasis on
surface area we found that surface charge makes larger contribution to
protein-solvent interaction than area does. These findings are
consistent with water being a protic solvent that prefers anions over
cations and show that folding into a native state is to optimize
surface negative charge rather to minimize hydrophobic surface
area. They suggest that the optimization of protein solvation through
natural selection is achieved by (1) universal enrichment of surface
negative charge, (2) increased surface areas for hydrophilic atoms for
better hydrogen bonding with solvent, and (3) higher concave-convex
ratio for hydrophobic atoms for better VDW attraction with solvent.
{\small \bibliography{surface} \bibliographystyle{unsrt} }

\begin{thebibliography}{10}

\bibitem{Kendrew1960}
J.~C. {Kendrew}, R.~E. Dickerson, B.~E. Strandberg, R.~G. Hart, D.~R. Davies,
  D.~C. Phillips, and V.~C. Shor.
\newblock {Structure of Myoglobin: A Three-Dimensional Fourier Synthesis at
  2\AA~Resolution}.
\newblock {\em Nature}, 185:422--427, February 1960.

\bibitem{Lee1971379}
B.~Lee and F.~M. Richards.
\newblock The interpretation of protein structures: Estimation of static
  accessibility.
\newblock {\em Journal of Molecular Biology}, 55(3):379--400, 1971.

\bibitem{ASAdefinition}
Robert~B. Hermann.
\newblock Theory of hydrophobic bonding. {II}. correlation of hydrocarbon
  solubility in water with solvent cavity surface area.
\newblock {\em The Journal of Physical Chemistry}, 76(19):2754--2759, 1972.

\bibitem{FMRichards1977}
F.~M. Richards.
\newblock Areas, volumes, packing, and protein structure.
\newblock {\em Annual Review of Biophysics and Bioengineering}, 6(1):151--176,
  1977.
\newblock PMID: 326146.

\bibitem{Chothia1974}
C.~Chothia.
\newblock Hydrophobic bonding and accessible surface area in proteins.
\newblock {\em Nature}, 248:338--339, 1974.

\bibitem{Tanford1974}
J.~A. Reynolds, D.~B. Gilbert, and C.~Tanford.
\newblock Empirical correlation between hydrophobic free energy and aqueous
  cavity surface area.
\newblock {\em Proc. Natl. Acad. Sci. USA}, 71(8):2925--2927, 1974.

\bibitem{Eisenberg1986}
D.~Eisenberg and A.~D. Andrew D.~McLachlan.
\newblock Solvation energy in protein folding and binding.
\newblock {\em Nature}, 319:199--203, 1986.

\bibitem{OOI1987}
T.~Ooi, M.~Oobatake, G.~Nmmethy, and H.~A. Scheraga.
\newblock Accessible surface areas as a measure of the thermodynamic parameters
  of hydration of peptides.
\newblock {\em Proc. Natl. Acad. Sci. USA}, 84(10):3086--3090, 1987.

\bibitem{Sharp106}
K.~A. Sharp, A.~Nicholls, R.~F. Fine, and B.~Honig.
\newblock Reconciling the magnitude of the microscopic and macroscopic
  hydrophobic effects.
\newblock {\em Science}, 252(5002):106--109, 1991.

\bibitem{charmm2009}
B.~R. Brooks, C.~L. Brooks, A.~D. Mackerell, L.~Nilsson, R.~J. Petrella,
  B.~Roux, Y.~Won, G.~Archontis, C.~Bartels, S.~Boresch, A.~Caflisch, L.~Caves,
  Q.~Cui, A.~R. Dinner, M.~Feig, S.~Fischer, J.~Gao, M.~Hodoscek, W.~Im,
  K.~Kuczera, T.~Lazaridis, J.~Ma, V.~Ovchinnikov, E.~Paci, R.~W. Pastor, C.~B.
  Post, J.~Z. Pu, M.~Schaefer, B.~Tidor, R.~M. Venable, H.~L. Woodcock, X.~Wu,
  W.~Yang, D.~M. York, and M.~Karplus.
\newblock Charmm: The biomolecular simulation program.
\newblock {\em Journal of Computational Chemistry}, 30(10):1545--1614, 2009.

\bibitem{CONNOLLY:PQMS}
M.~L. Connolly.
\newblock The molecular surface package.
\newblock {\em Journal of Molecular Graphics}, 11(2):139--141, 1993.

\bibitem{MSMS}
M.~F. Sanner, A.~J. Olson, and J.~Spehner.
\newblock Reduced surface: An efficient way to compute molecular surfaces.
\newblock {\em Biopolymers}, 38(3):305--320, 1996.

\bibitem{CHOTHIA19761}
C.~Chothia.
\newblock The nature of the accessible and buried surfaces in proteins.
\newblock {\em Journal of Molecular Biology}, 105(1):1--12, 1976.

\bibitem{Ashbaugh1999}
H.~S. Ashbaugh, E.~W. Kaler, and M.~E. Paulaitis.
\newblock A “universal” surface area correlation for molecular hydrophobic
  phenomena.
\newblock {\em Journal of the American Chemical Society}, 121(39):9243--9244,
  1999.

\bibitem{Singh2006145}
Y.~Hemajit Singh, M.~Michael Gromiha, Akinori Sarai, and Shandar Ahmad.
\newblock Atom-wise statistics and prediction of solvent accessibility in
  proteins.
\newblock {\em Biophysical Chemistry}, 124(2):145 -- 154, 2006.

\bibitem{Kapcha2014484}
Lauren~H. Kapcha and Peter~J. Rossky.
\newblock A simple atomic-level hydrophobicity scale reveals protein
  interfacial structure.
\newblock {\em Journal of Molecular Biology}, 426(2):484--498, 2014.

\bibitem{Janin1979}
J.~Janin.
\newblock {Surface and Inside Volumes in Globular Proteins}.
\newblock {\em Nature}, 277:491--492, February 1979.

\bibitem{KYTE1982105}
Jack Kyte and Russell~F. Doolittle.
\newblock A simple method for displaying the hydropathic character of a
  protein.
\newblock {\em Journal of Molecular Biology}, 157(1):105--132, 1982.

\bibitem{EWscale1984}
D.~Eisenberg, R.~M. Weiss, and T.~C. Terwilliger.
\newblock The hydrophobic moment detects periodicity in protein hydrophobicity.
\newblock {\em Proc. Natl. Acad. Sci. USA}, 81(1):140--144, 1984.

\bibitem{GES1986}
D.~M. Engelman, T.~A. Steitz, and Goldman. A.
\newblock Identifying nonpolar transbilayer helices in amino acid sequences of
  membrane proteins.
\newblock {\em Annual Review of Biophysics and Biophysical Chemistry},
  15(1):321--353, 1986.
\newblock PMID: 3521657.

\bibitem{sesProtLigand}
Lincong Wang.
\newblock The constributions of surface charge and geometry to protein-ligand
  interaction.
\newblock {\em Manuscript in preparation}.

\bibitem{Kauzmann1987}
W.~{Kauzmann}.
\newblock {Thermodynamics of unfolding}.
\newblock {\em Nature}, 325:763--764, February 1987.

\bibitem{Dill1990}
K.~A. Dill.
\newblock Dominant forces in protein folding.
\newblock {\em Biochemistry}, 29(31):7133--7155, 1990.
\newblock PMID: 2207096.

\bibitem{Baldwin2014}
R.~L. Baldwin.
\newblock Dynamic hydration shell restores kauzmann’s 1959 explanation of how
  the hydrophobic factor drives protein folding.
\newblock {\em Proc. Natl. Acad. Sci. USA}, 111(36):13052--13056, 2014.

\bibitem{Onsager1936}
Lars Onsager.
\newblock Electric moments of molecules in liquids.
\newblock {\em Journal of the American Chemical Society}, 58(8):1486--1493,
  1936.

\bibitem{CKmodel1957}
C.~Tanford and J.~G. Kirkwood.
\newblock Theory of protein titration curves. i. general equations for
  impenetrable spheres.
\newblock {\em Journal of the American Chemical Society}, 79(20):5333--5339,
  1957.

\bibitem{WARSHEL1976227}
A.~Warshel and M.~Levitt.
\newblock Theoretical studies of enzymic reactions: Dielectric, electrostatic
  and steric stabilization of the carbonium ion in the reaction of lysozyme.
\newblock {\em Journal of Molecular Biology}, 103(2):227--249, 1976.

\bibitem{Warshel20061647}
Arieh Warshel, Pankaz~K. Sharma, Mitsunori Kato, and William~W. Parson.
\newblock Modeling electrostatic effects in proteins.
\newblock {\em Biochimica et Biophysica Acta (BBA) - Proteins and Proteomics},
  1764(11):1647--1676, 2006.

\bibitem{solvationModel2009}
Aleksandr~V. Marenich, Christopher~J. Cramer, and Donald~G. Truhlar.
\newblock Universal solvation model based on solute electron density and on a
  continuum model of the solvent defined by the bulk dielectric constant and
  atomic surface tensions.
\newblock {\em The Journal of Physical Chemistry B}, 113(18):6378--6396, 2009.
\newblock PMID: 19366259.

\bibitem{PROT:PROT21715}
Pascal Benkert, Silvio C.~E. Tosatto, and Dietmar Schomburg.
\newblock Qmean: A comprehensive scoring function for model quality assessment.
\newblock {\em Proteins: Structure, Function, and Bioinformatics},
  71(1):261--277, 2008.

\bibitem{Kota15082011}
Pradeep Kota, Feng Ding, Srinivas Ramachandran, and Nikolay~V. Dokholyan.
\newblock Gaia: automated quality assessment of protein structure models.
\newblock {\em Bioinformatics}, 27(16):2209--2215, 2011.

\bibitem{cnsVersion1dot2}
A.~T. Brunger.
\newblock {Version 1.2 of the Crystallography and {NMR} System}.
\newblock {\em Nature Protocol}, 2:2728--2733, February 2007.

\bibitem{reduce1999}
J.Michael Word, Simon~C. Lovell, Jane~S. Richardson, and David~C. Richardson.
\newblock Asparagine and glutamine: using hydrogen atom contacts in the choice
  of side-chain amide orientation1.
\newblock {\em Journal of Molecular Biology}, 285(4):1735--1747, 1999.

\bibitem{Perutz1187}
M.~F. Perutz.
\newblock Electrostatic effects in proteins.
\newblock {\em Science}, 201(4362):1187--1191, 1978.

\bibitem{KADill2008}
D.~L. Mobley, A.~E. Barber~II, C.~J. Fennell, , and K.~A. Dill.
\newblock Charge asymmetries in hydration of polar solutes.
\newblock {\em The Journal of Physical Chemistry B}, 112(8):2405--2414, 2008.
\newblock PMID: 18251538.

\bibitem{Simonson1995}
T.~Simonson and D.~Perahia.
\newblock Internal and interfacial dielectric properties of cytochrome c from
  molecular dynamics in aqueous solution.
\newblock {\em Proc. Natl. Acad. Sci. USA}, 92(4):1082--1086, 1995.

\bibitem{Tanford1979}
C.~Tanford.
\newblock Interfacial free energy and the hydrophobic effect.
\newblock {\em Proc. Natl. Acad. Sci. USA}, 76(9):4175--4176, 1979.

\bibitem{Zhou2013}
X.~Pang and H.~X. Zhou.
\newblock {Poisson-Boltzmann} calculations: van der {Waals} or molecular
  surface?
\newblock {\em Commun Comput Phys}, 13(1):1--12, 2013.

\bibitem{Pettitt2014}
R.~C. Harris, B., and B.~M. Pettitt.
\newblock Effects of geometry and chemistry on hydrophobic solvation.
\newblock {\em Proc. Natl. Acad. Sci. USA}, 111(41):14681--14686, 2014.

\bibitem{Rose834}
G.~D. Rose, A.~R. Geselowitz, G.~J. Lesser, R.~H. Lee, and M.~H. Zehfus.
\newblock Hydrophobicity of amino acid residues in globular proteins.
\newblock {\em Science}, 229(4716):834--838, 1985.

\end{thebibliography}
\newpage
\section*{Supplementary Information}

\subsection*{S1: The comparison of the surfaces by MSMS, PyMOL and our SES program}
The structure used for the comparison is a peptide (min1.pdb)
downloaded from an Amber tutorials website
(http://ambermd.org/tutorials/basic/tutorial3/section3.htm). The MSMS
SES surface is generated by setting a probe radius=1.2\AA,
density=100.0 and high density=200.0. During the surface computation
the atomic radii for the following three atoms, 66, 69 and 97, are
increased by $0.1$\AA. As indicated by the four arrows in Fig.~S1a the
regions where the PAA areas interact with each other are relatively
rough. In the contrast the four corresponding regions (colored in
violet) in our SES surface (Fig.~S1b) are as smooth as the rest. As a
reference the surface (Fig.~S1c) generated by a popular molecular
visualization program PyMOL lacks the details especially for the four
intersecting PAA regions.
\begin{figure} [htp!]
\centering \subfloat[MSMS]{\label{fig:msmsSES}
  \includegraphics[width=0.75\textwidth]{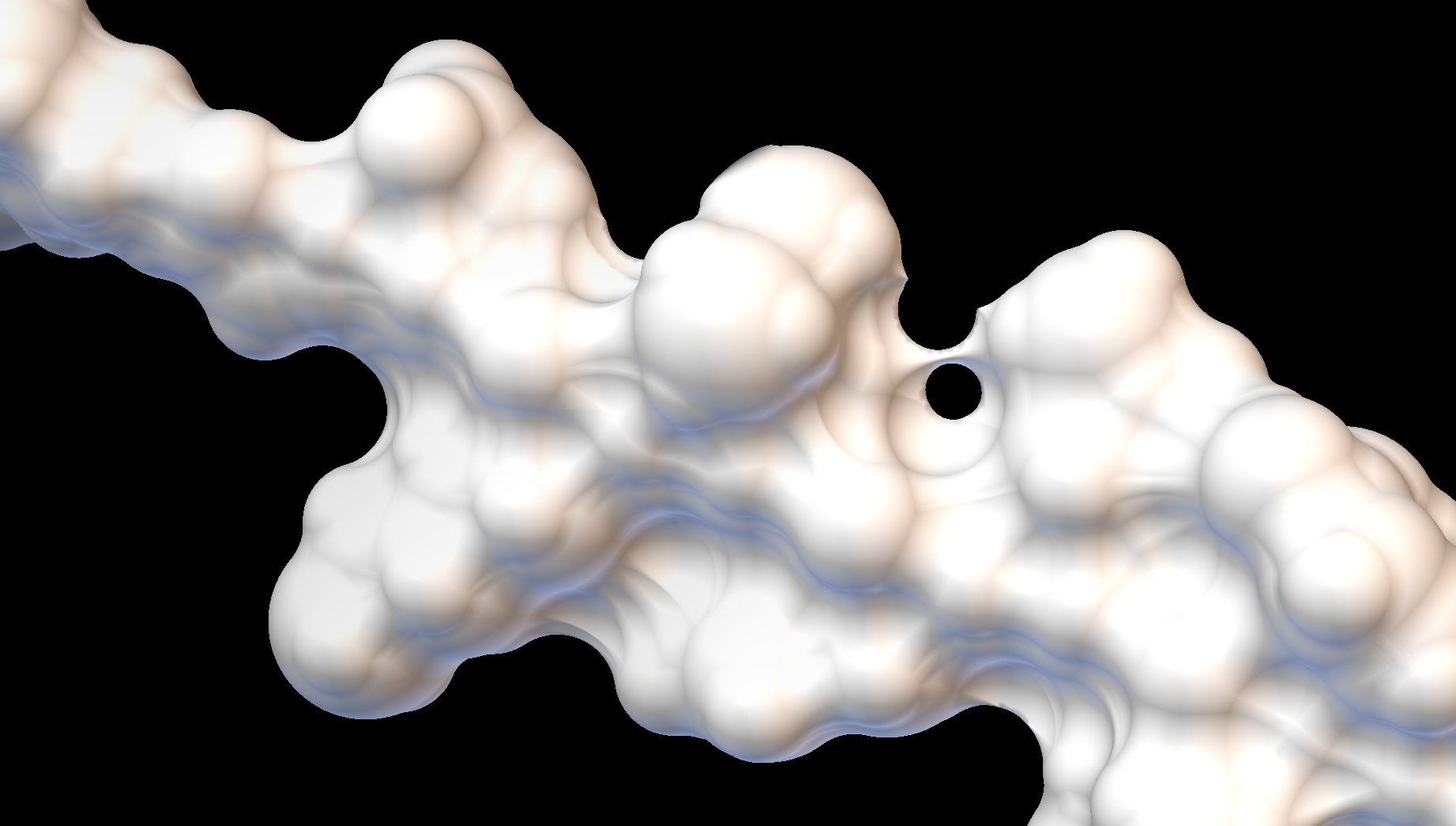}}\\ \subfloat[Our
  SES Program]{\label{fig:ourSES}
  \includegraphics[width=0.75\textwidth]{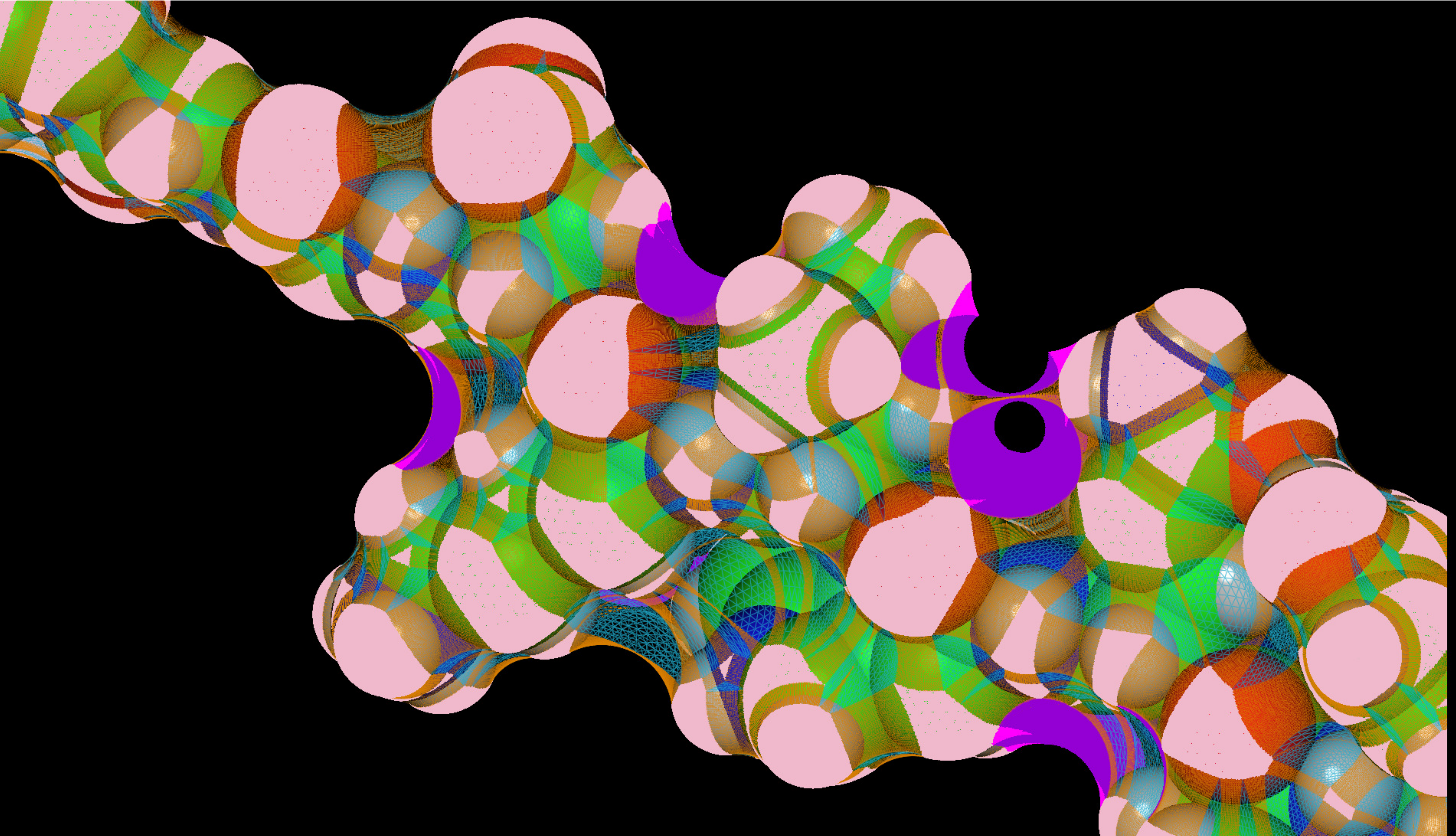}}\\ \subfloat[PyMOL]{\label{fig:pymol}
  \includegraphics[width=0.75\textwidth]{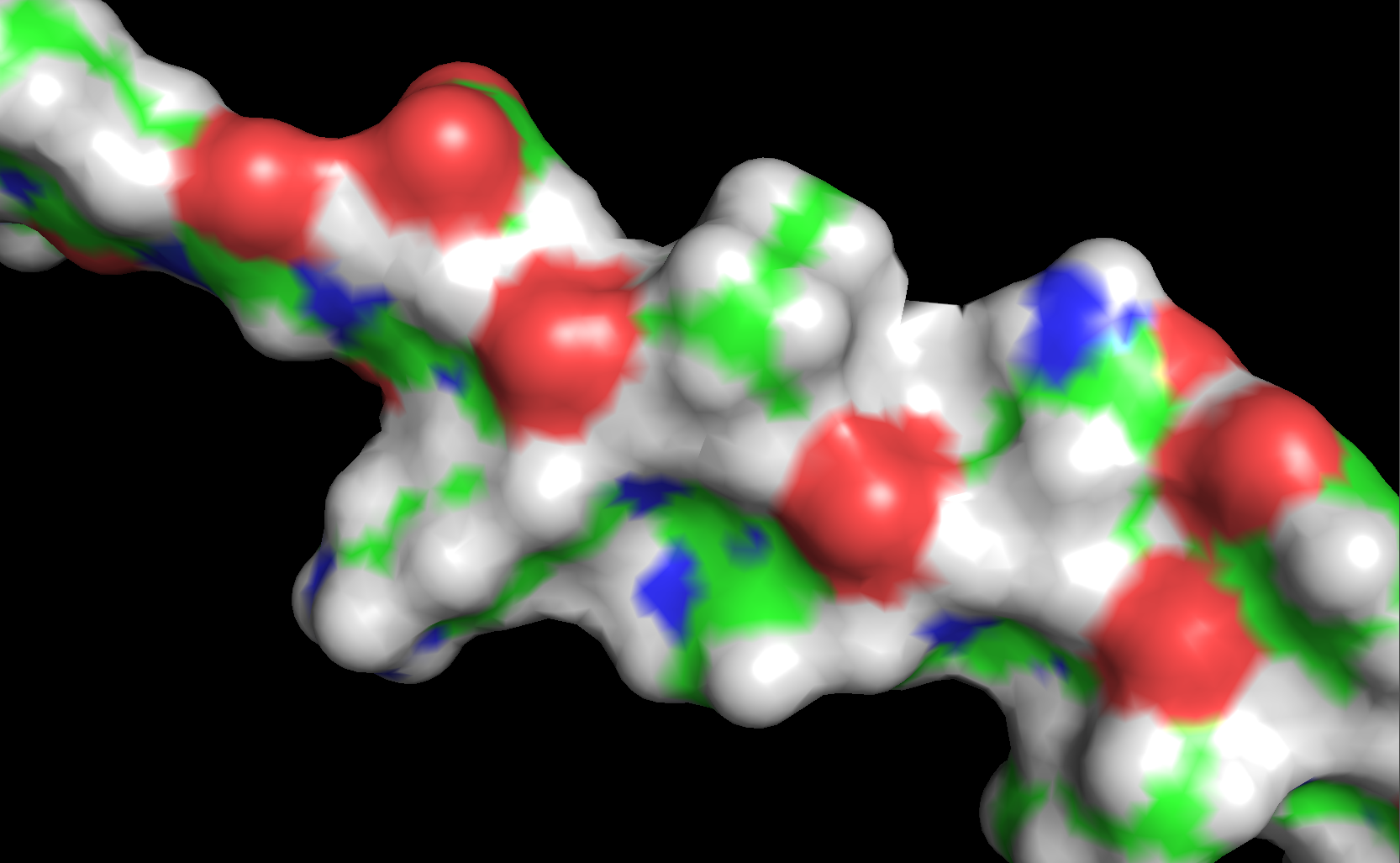}}
\caption* {\textbf{Figure S1:~The surfaces by MSMS, our program and PyMOL.}  } 

\end{figure}

\end{document}